\documentclass[preprint,preprintnumbers,amsmath,amssdoes notymb]{revtex4}
\usepackage{amsmath}
\usepackage{amssymb}
\usepackage{bm}
\usepackage{dcolumn}
\usepackage{graphicx}
\usepackage{color}

\begin{document}

\title{A quantum Monte Carlo study of the ground state chromium dimer }

\author{Kenta Hongo$^{1,2,3}$}
\author{Ryo Maezono$^1$}
\address{$^1$School of Information Science, Japan Advanced Institute of 
Science and Technology, Asahidai 1-1, Nomi, Ishikawa, 923-1292, Japan}
\address{$^2$Department of Chemistry and Chemical Biology, Harvard 
University, Cambridge, Massachusetts, 02138}
\address{$^3$JSPS Postdoctoral Fellow for Research Abroad}

\date{\today}

\begin{abstract}
We report  variational and diffusion quantum Monte Carlo (VMC and DMC) 
studies of the binding curve of the ground-state chromium dimer. 
We employed various single determinant (SD) or multi-determinant (MD) wavefunctions 
multiplied by a Jastrow fuctor as a trial/guiding wavefunction. 
The molecular orbitals (MOs) in the SD were
calculated using restricted or unrestricted  
Hartree-Fock or density functional theory (DFT) calculations where 
five commonly-used local (SVWN5), semi-local (PW91PW91 and BLYP), 
and hybrid (B1LYP and B3LYP) functionals were examined. 
The MD expansions were obtained from the complete-active space SCF, 
generalized valence bond, and unrestricted configuration interaction methods. 
We also adopted the UB3LYP-MOs to construct the MD expansion (UB3LYP-MD) 
and optimized their coefficients at the VMC level. 
In addition to the wavefunction dependence, 
we investigated the time-step bias in the DMC calculation
and the effects of pseudopotentials and backflow transformation for the UB3LYP-SD case. 
Some of the VMC binding curves show a flat or quite shallow well bottom, 
which gets recovered deeper by DMC.
All the DMC binding curves have a minimum indicating a bound state, 
but the comparison of atomic and molecular energies 
gives rise to a negative binding energy for all the DMC 
as well as VMC calculations. 

\end{abstract}

\maketitle

\section{Introduction}
\label{section:introduction}
\par
The chromium dimer (Cr$_2$) has attracted a lot of attention as a prototype 
to understand the $d$-$d$ binding in both experimental
\cite{Michalopoulos_etal_1982,Bondybey_English_1983,Simard_etal_1998,Hilpert_Ruthardt_1987, Su_etal_1993, Casey_Leopold_1993, Moskovits_etal_1985} and 
theoretical studies \cite{McLean_Liu_1983,Richman_McCullough_1987,Walch_etal_1983,Wood_etal_1980,Goodgame_Goddard_1981, Goodgame_Goddard_1985,Anderson_etal_1994,Anderson_Roos_1995, Anderson_etal_1996, Celani_etal_2004, Mitrushenkov_Palmieri_1997,Angeli_etal_2002, Angeli_etal_2006, Scuseria_1991, Baushlicher_Partridge_1994, Meuller_etal_2009, Stoll_Werner_1996,Atha_Hillier_1982, Kok_Hall_1983, Vissher_etal_1994, Dachsel_etal_1999, Takahara_eta_1989, Moritz_etal_2005, Kurashige_Yanai_2009, Delley_etal_1983, Bernholc_Hozwarth_1983, Dunlap_1983, Baykara_etal_1984, Painter_1986, Edgecombe_Becke_1995, Cheng_Wang_1996, Desmarais_etal_2000, Yanagisawa_etal_2000, Barden_etal_2000, Valiev_etal_2003, Furche_Perdew_2006, Calaminici_etal_2007, Gutsev_Bauschlicher_2003, Thomas_etal_1999, Kitagawa_etal_2007, Shi-Ying_2008, Tsuchimochi_etal_2010}. 
The ground state is experimentally 
observed to be a singlet state, $^1\Sigma_g^+$\cite{Michalopoulos_etal_1982}, 
whereas the ground state of the constituent Cr atom $3d^5 4s^1$ ($^7S$) 
has the highest spin multiplicity in the $3d$ atoms. 
Recent spectroscopic experiments reported an 
equilibrium bond length ($R_e$) and binding energy ($D_e$) of 1.6788 \AA 
\cite{Bondybey_English_1983}  and 1.53 $\pm$ 0.05 eV\cite{Simard_etal_1998}, 
respectively (some older experiments reported $R_e$ of 1.68 \AA
\cite{Michalopoulos_etal_1982} and $D_e$ of 1.44 $\pm$ 0.05 eV 
\cite{Hilpert_Ruthardt_1987} and 1.42 $\pm$ 0.10 eV \cite{Su_etal_1993}). 
However, no theoretical study has provided a quantitatively satisfactory result 
for Cr$_2$ so far, since its chemical binding is rather complicated. 
\par
From the theoretical viewpoint, there are two extremely different pictures to 
understand the chemical binding in  Cr$_2$ qualitatively. The first one is based 
on elementary molecular orbital (MO) theories. In this framework, Cr$_2$ is 
treated as a closed shell (single-determinant) state, with all the bonding orbitals 
occupied ($1\sigma_g^2 2\sigma_g^2 1\pi_u^4 1\delta_g^4$, arising from the 
$3d$ and $4s$ orbitals). 
Cr$_2$ is therefore interpreted as having a sextuple bond, 
which is the highest multiple bond in any diatomic molecule. This is naively 
consistent with the observed short bond length ($\approx$ 1.68 \AA)
\cite{Michalopoulos_etal_1982, Bondybey_English_1983}, whereas the 
experimental $D_e$ ($\approx$ 1.5 eV) is rather small in the context of multiple 
bonds, which is smaller than that of singly bonded Cu$_2$ ($\approx$ 2.0 eV). 
This fact may imply that an elementary one-electron approximation picture is invalid 
for Cr$_2$ and hence ``non-dynamical" (static) correlation effects are important. 
Indeed, restricted Hartree-Fock (RHF) does not only give an incorrect dissociation 
behavior, but also gives rise to a ground state energy of Cr$_2$ far above that 
of the two isolated atoms (about 20 eV higher)\cite{Goodgame_Goddard_1981}. 
\par
The second picture,  which is the extreme opposite to the first one, emphasizes 
the relative stability of high-spin atomic states\cite{Goodgame_Goddard_1981}. 
This is the molecular analogue of antiferromagnetism and can be treated at the 
crudest level of theory by the unrestricted (or broken symmetry) Hartree-Fock 
(UHF) theory using a spin unrestricted single-determinant of symmetry broken 
molecular spin orbitals\cite{Takahara_eta_1989,Kitagawa_etal_2007}. 
The method can approximately deal with the static correlation effects. 
Up and down-spin electrons are antiferromagnetically 
localized on each of the Cr atoms, 
where the charge density has a $D_{\infty h}$ symmetry 
but the spin density has a $C_{\infty v}$ symmetry. 
The wavefunction represents the nearest possible single-determinant approximation 
to the $^1\Sigma_g^+$ state, but it is not an eigenfunction of total spin, leading 
to spin contamination.
This can be easily remedied by a properly symmetrized 
multi-determinant expansion of the UHF wavefunction, 
{\it i.e.}, the generalized 
valence bond (GVB) method. However, the GVB method gives rise to a very 
weakly bound molecule ($D_e = 0.35$ eV) with a very long bond length ($R_e = 
3.1$ \AA)\cite{Goodgame_Goddard_1981}. Although a complete active space 
(CAS) self-consistent field (SCF) method can also consider the static correlation, 
a typical CASSCF with a 12 orbital active space and 12 active electrons, 
CASSCF(12,12), provides a poor result of $R_e$ and $D_e$, similar to GVB
\cite{Kitagawa_etal_2007}. 
These disagreements with experiments indicate dynamical correlation effects are 
also important for achieving quantitatively satisfactory results. In summary, the 
chemical binding in Cr$_2$ involves a highly complicated blend of $4s$-$4s$ 
and $3d$-$3d$ interactions with antiferromagnetic coupling.
\par
To understand such a complicated binding mechanism, a large number of 
{\it ab initio} studies have been performed based on either
traditional MO theories\cite{McLean_Liu_1983, 
Richman_McCullough_1987, Walch_etal_1983, Wood_etal_1980, 
Goodgame_Goddard_1981, Goodgame_Goddard_1985, Anderson_etal_1994, 
Anderson_Roos_1995, Anderson_etal_1996, Celani_etal_2004, 
Mitrushenkov_Palmieri_1997, Angeli_etal_2002, Angeli_etal_2006, Scuseria_1991, 
Baushlicher_Partridge_1994, Meuller_etal_2009, 
Stoll_Werner_1996,Atha_Hillier_1982, Kok_Hall_1983, Vissher_etal_1994, 
Dachsel_etal_1999, Takahara_eta_1989, Moritz_etal_2005, 
Kurashige_Yanai_2009} 
or density functional theories (DFT)
\cite{Delley_etal_1983, Bernholc_Hozwarth_1983, Dunlap_1983, 
Baykara_etal_1984, Painter_1986, Edgecombe_Becke_1995, Cheng_Wang_1996, 
Desmarais_etal_2000, Yanagisawa_etal_2000, Barden_etal_2000, 
Valiev_etal_2003, Furche_Perdew_2006, Calaminici_etal_2007, 
Gutsev_Bauschlicher_2003, Thomas_etal_1999, Kitagawa_etal_2007, Shi-Ying_2008, Tsuchimochi_etal_2010}.  
In MO, both single- and multi-reference theories were studied.
Within the single-reference theory, the coupled-cluster approach with single 
and double excitations 
including triples noniteratively, CCSD(T), is one of the most 
accurate methods. 
Restricted CCSD(T) calculations give a very weak
binding ($D_e = 0.38 $ eV) with a short bond length ($R_e = 1.60$ 
\AA), whereas unrestricted CCSD(T) calculations provide a better 
binding energy ($D_e = 0.89$ eV), but with a longer bond length ($R_e = 
2.54$ \AA). 
In multi-reference theories, a typical reference is a CASSCF(12,12) calculation 
which is responsible for the static correlation. 
Using the CAS reference, dynamical correlation can be taken into account by 
multi-reference configuration interaction (MRCI)\cite{Dachsel_etal_1999}, 
coupled cluster (MRCC)\cite{Meuller_etal_2009}, or  second-order perturbation 
(CASPT2) methods\cite{Anderson_etal_1994}. 
Although these multi-reference 
methods give a better $R_e$ than single-reference methods, 
there is still room to improve further the accuracy of $D_e$.

\par
In DFT, various exchange-correlation (XC) functionals are available such as 
the local (spin) density approximation (LDA/LSDA), the generalized gradient 
approximation (GGA), as well as the hybrid XC functionals.
Both restricted LDA (RLDA) and the unrestricted formalism (ULDA) give rise to  
overbinding, {\it i.e.}, too short $R_e$ and too large $D_e$
\cite{Delley_etal_1983,Bernholc_Hozwarth_1983,Dunlap_1983, 
Baykara_etal_1984,Painter_1986}, 
which is a well-known failure of LDA.
Both restricted and unrestricted GGA calculations\cite{Edgecombe_Becke_1995, 
Thomas_etal_1999, Desmarais_etal_2000, 
Yanagisawa_etal_2000,Barden_etal_2000,Valiev_etal_2003, 
Gutsev_Bauschlicher_2003, Furche_Perdew_2006, Calaminici_etal_2007} 
generally improve the LDA discrepancy, 
but it is difficult to choose the ``best" GGA functional because one may give a better $R_e$, 
but it may give a worse $D_e$, and vice versa. 
Restricted B3LYP, 
though it is popular for covalent molecules, gives rise to an unbound 
molecular state\cite{Edgecombe_Becke_1995,Kitagawa_etal_2007}. 
Although unrestricted B3LYP reproduces a bound state, 
it provides a smaller $D_e$ of $\approx 1.0$ eV with much larger $R_e$ of 
$\approx 2.5 $ \AA
\cite{Edgecombe_Becke_1995,Kitagawa_etal_2007}. 
These results may imply that the difficulties for the binding of Cr$_2$
would originate from a delicate balance between exchange and correlation in DFT
for chemical binding of Cr$_2$.

\par
Quantum Monte Carlo (QMC) methods\cite{Hammond_Lester_Reynolds_1994, 
Foulkes_Mitas_Needs_Rajagopal_2001}  are one of the most accurate 
techniques in state-of-the-art {\it ab-initio} calculations for quantitative 
descriptions of electronic structures.  
There are two typical QMC calculations, {\it i.e.}, 
variational and diffusion Monte Carlo (VMC and DMC) methods. 
VMC is not usually 
accurate enough since its result strongly depends on the correlated trial 
wavefunction adopted. DMC is a technique for numerically solving the many-
electron Sch\"{o}dinger equation for stationary states using imaginary time 
evolution. 
The fixed-node approximation is usually assumed to maintain the 
fermionic anti-symmetry in DMC. 
Although the fixed-node DMC can accurately evaluate the ground state energy of 
many atoms and molecules using only the trial node from a single determinant (SD), 
it sometimes fails, especially for
``near-degeneracy" systems such as the Be atom. 
This implies that the fixed-node DMC method can work well for the dynamic 
correlation, but not for the static correlation which should be included at the 
stage of choosing the fixed-node trial wavefunction. The Cr$_2$ molecule is 
also considered as a near-degeneracy molecular system and hence a good 
challenge to QMC. 

\par
In this study we performed VMC and DMC calculations of Cr$_2$ with 
several choice of trial wavefunctions.
Variety of the XC functionals are examined to construct orbital functions including
HF,  SVWN LDA, PW91 and BLYP GGA, B1LYP and B3LYP hybrid functionals,
with both restricted and unrestricted treatments.
In addition to a single-determinant form of the many-body wavefunction,
we also tried several multi-determinant (MD) forms. 
For the orbital part we also introduced the backflow transformation \cite{FEY56,DRU06,RIO06}.
Performance of the XC functionals are examined based on the variational principle 
with respect to the fixed node \cite{Hammond_Lester_Reynolds_1994,Foulkes_Mitas_Needs_Rajagopal_2001}.
The chemical binding of the ground-state Cr$_2$ may be examined in two ways: 
(i) use of the binding curve and (ii) a comparison of atomic and molecular energies. 

\par
The present paper is organized as follows: Section II describes the present 
computational methods. Section III   provides numerical results and discussion. 
Section IV summarizes the present study. 

\par
\section{Computational methods}
We replaced the inner Neon core (10 core electrons) of the Cr atom 
with a small core norm-conserving pseudopotential which is constructed 
from Dirac-Fock atomic solutions.
More specifically we employed the Lee-Needs (LN) soft pseudopotential 
\cite{LEE00} using Troullier-Martins construction.
To check the dependence on the potential 
we also used a small core Burkatzki-Filippi-Dolg (BFD) pseudopotential~\cite{BUR08}.
An electronic many-body wavefunction is
composed of anti-symmetrized products of orbital functions
which is generated from DFT or HF.
Various XC functionals give different orbitals
from which different structures of the nodal surface \cite{CEP91}
of the many-body wavefunction 
(the surface in configuration space on which the wavefunction 
is zero and across which it changes sign) 
are constructed.
For different nodal structures, we can exploit the variational principle to see
which choice is better, comparing the ground state energy\cite{WAG03}.
In this study we tested several combinations of
(i) the choice of XC, (ii) spin restricted/unrestricted treatment of orbitals, (iii) SD/MD, 
(iv) choice of pseudopotentials, and (v) with/without
backflow degrees of freedom\cite{FEY56}.
\par
In VMC the ground state
energy is evaluated as the expectation value of the
Hamiltonian $\hat{H}$ with a many-body trial wavefunction, $\Psi$,
\begin{equation}
\label{eq:vmc_energy}
E = \frac{\int \Psi^* \hat{H} \Psi \, d{\bf R}}{\int \Psi^* \Psi \,
d{\bf R}} = \frac{\int |\Psi|^2 \Psi^{-1} \hat{H} \Psi \, d{\bf
R}}{\int |\Psi|^2 \, d{\bf R}} ,
\end{equation}
where $\bf R = \left({\bf r}_{1},...,{\bf r}_{N}\right)$ denotes
an electronic configuration of valence electrons in a molecule ($N=28$ with
14 up and down spins, respectively, for Cr$_2$),
and the energy has been written as an average of the
local energy, $E_L \left( {\bf R} \right) = \Psi^{-1} \hat{H} \Psi$, over the probability
distribution $p({\bf R}) = |\Psi|^2/\int |\Psi|^2 \, d{\bf R}$.  
The energy expectation value is evaluated from Monte Carlo integration,
using the Metropolis algorithm to generate electronic configurations
distributed according to $p({\bf R})$.  
\par
In the DMC method the ground-state component of a trial wavefunction 
which overlaps with the exact one
is projected out by evolving an ensemble of electronic configurations
using the imaginary-time Schr\"odinger equation.  
Attempts to carry
out this procedure exactly result in a ``fermion sign problem'', which
is circumvented by constraining the nodal surface of the wavefunction 
to equal that of the trial wavefunction.  
The DMC energy calculated with this fixed-node constraint
is equal or higher than the exact ground-state energy, 
and becomes equal to the exact one if and only if the fixed nodal surface is exact.  

\par
We used the many-body wavefunction taking the form,
\begin{equation}
\label{10.02.04.math01}
\Psi \left( {\bf R} \right) = e^{J\left({\bf R} \right)}
\cdot F_{\rm AS} \left( {{\bf r}_1 , \cdots ,{\bf r}_N } \right) \ ,
\end{equation}
where $F_{\rm AS} \left( {{\bf r}_1 , \cdots ,{\bf r}_N } \right)$
is the anti-symmetrized products of orbital functions
$\left\{\psi_i^{\sigma}\left({\bf r}_j\right)\right\}$
such as a Slater determinant. 
In this study the function is expanded as
\begin{equation}
\label{F_function}
F_{\rm AS} \left( {{\bf r}_1 , \cdots ,{\bf r}_N } \right) 
= \sum\limits_{i = 0}^{N_{MD}^ \uparrow} 
\sum\limits_{j = 0}^{N_{MD}^ \downarrow} 
c_{ij} \cdot D_i^ \uparrow \left(\tilde{\bf r}_1, \cdots ,
\tilde{ \bf  r}_{{N \mathord{\left/
{\vphantom {N 2}} \right.\kern-\nulldelimiterspace} 2}}  \right)
D_j^ \downarrow  \left( \tilde{\bf r}_{{N \mathord{\left/
{\vphantom {N 2}} \right.\kern-\nulldelimiterspace} 2} + 1}
 , \cdots ,\tilde{\bf r}_N\right) \ ,
\end{equation}
by determinants with coefficients $c_{ij}$.
$D_0^{\sigma = \uparrow,\downarrow}  \left( \tilde{\bf r}_1
 , \cdots ,\tilde{\bf r}_{N/2}\right)$ is the ground-state 
single Slater determinant formed only by the occupied orbitals for each spin,
and $D_{j \ne 0}^\sigma   = \hat P_{n \to m}^{\left(j\right)} \cdot D_0^\sigma$
corresponds to excited state configurations, 
where $\hat P_{n \to m}^{\left(j\right)}$ 
denotes the excitation from the occupied state
$\psi_n^{\sigma}\left({\bf r}_j\right)$ into the virtual state
$\psi_m^{\sigma}\left({\bf r}_j\right)$.
In Eq.~\!\eqref{F_function}
arguments with a tilde
$\tilde{\bf r}_i ={\bf r}_i  + {\bf \xi} _i \left(  {\bf r}_1,\cdots, {\bf r}_N\right)$,
denote the backflow shift \cite{FEY56}.
$N_{\rm MD}^\sigma=0$ corresponds to SD, and
$c_{ij}=c_i \cdot \delta_{ij}, D_i^\uparrow = D_i^\downarrow$ to
a spin restricted wavefunction.
The orbital functions, 
$\left\{\psi_i^{\sigma}\left({\bf r}_j\right)\right\}$,
were expanded with a contracted Gaussian basis set
(17s18p15d6f)/$\left[ {\rm 8s8p7d3f} \right]$.
Gaussian03\cite{Gaussian03} was used for SCF calculations, 
while we used CASINO\ ver.3.0\cite{CASINO} for QMC calculations.
Some calculations (HF and GVB) were carried out 
using GAMESS\cite{GAM93} for SCF and QWalk\cite{QWA09} for QMC.
We attempted to use another many-body wavefunction form, Pfaffian \cite{BAJ06}, which 
is available in QWalk.
However an optimization procedure for the off-diagonal elements of the Pfaffian did not work well and then we could not obtain reliable results, so not reported here in detail.
\par
As for the Jastrow factor $e^{J\left(\bf R\right)}$\cite{JAS55},
the function $J\left(\bf R\right)$ is given as 
\cite{DRU04},
\begin{equation}
J\left(\bf R\right)
= \sum\limits_{i = 1}^{N - 1} {\sum\limits_{j = i + 1}^N {u\left( {r_{ij} } \right)}  
+ \sum\limits_{I = 1}^{N_{\rm ions} } {\sum\limits_{i = 1}^N {\chi _I \left( {r_{iI} } 
\right)}  
+ \sum\limits_{I = 1}^{N_{\rm ions} } {\sum\limits_{i = 1}^{N - 1} 
{\sum\limits_{j = i + 1}^N {f_I } } } } } 
\left( {r_{iI} ,r_{jI} ,r_{ij} } \right)  ,
\label{10.02.04.math02}
\end{equation}
where suffices $i,j$ and $I$ specify electronic 
and ionic positions, respectively.
Each term, $u$, $\chi$, and $f$, takes into account
the dynamical correlation due to electron-electron, 
electron-nucleus, and electron-electron-nucleus
coalescence, respectively.
Cutoff lengths are introduced to make
each term quadratically fall off to zero at the radius.
Specific forms used in each QMC code are described in the appendix.
The electron-electron cusp condition \cite{KAT57}
is imposed in the $u$-term.
Variational parameters in $J\left(\bf R\right)$ are 
designed to be able to include spin polarized case and 
are optimized by VMC individually at each bond length with fixed cutoff lengths.
Those parameters were optimized by the variance minimization \cite{DRU05}
as well as the energy minimization \cite{UMR05} procedures.
The backflow shift \cite{FEY56},
${\bf \xi} _i \left( {\left\{ {{\bf r}_i } \right\}} \right)
={\bf \xi} _i^{ee} \left( {\left\{ {{\bf r}_{ij} } \right\}} \right)
+{\bf \xi}_i^{eN} \left( {\left\{ { {\bf r}_{iI} } \right\}} \right)$,
is introduced to modify the nodal surface variationally.
Each term, $ee$ and $eN$, is expanded with the power of inter-particle
distances (up to 8th order in this study)
with proper cutoff radii which are fixed 
in the same values as those in $u$ and $\chi$ in our Jastrow factor
\cite{DRU06,RIO06}.
The parameters in the backflow were optimized by
the filtered reweighted variance minimization scheme \cite{DRU05}
allowing spin polarized degrees of freedom.
\par
For the singlet Cr$_2$ ground state\cite{Michalopoulos_etal_1982,Bondybey_English_1983,Simard_etal_1998,Hilpert_Ruthardt_1987, Su_etal_1993, Casey_Leopold_1993, Moskovits_etal_1985}
the spin restricted SCF treatment seems not to give a proper description
of localized spin polarization on atomic sites \cite{SZA89}.
Even using the restricted nodal surface, however, 
the DMC projection can still recover the proper localized spin polarization.
Indeed we found the RHF nodal 
surface gives a variationally better description than the UHF one.
We therefore investigated spin restricted methods as a possibility
as well as unrestricted ones.
In order to generate the nodal surface, we employed five commonly-used XC functionals, 
SVWN5(LDA), PW91PW91 and BLYP (GGA), B1LYP and B3LYP(hybrid),
shown in Table. \ref{xc}.

\par
For DMC we started statistical accumulations of 2,000 walkers
after equilibration of 1,000 steps 
with $\delta t$= 0.01[a.u.], for which we have 
confirmed that the time step bias is within the statistical noise
considered here.
The present non-local pseudopotentials were evaluated by
the T-move scheme \cite{CAS06} which is devised to reduce
the instability and bias due to the locality approximation \cite{MIT91}.

\section{Results and Discussion} 
\subsection{Single determinant calculations}
\label{Single determinant calculations}
\begin{figure}[h]
\centering
\includegraphics[width=3truein,clip]{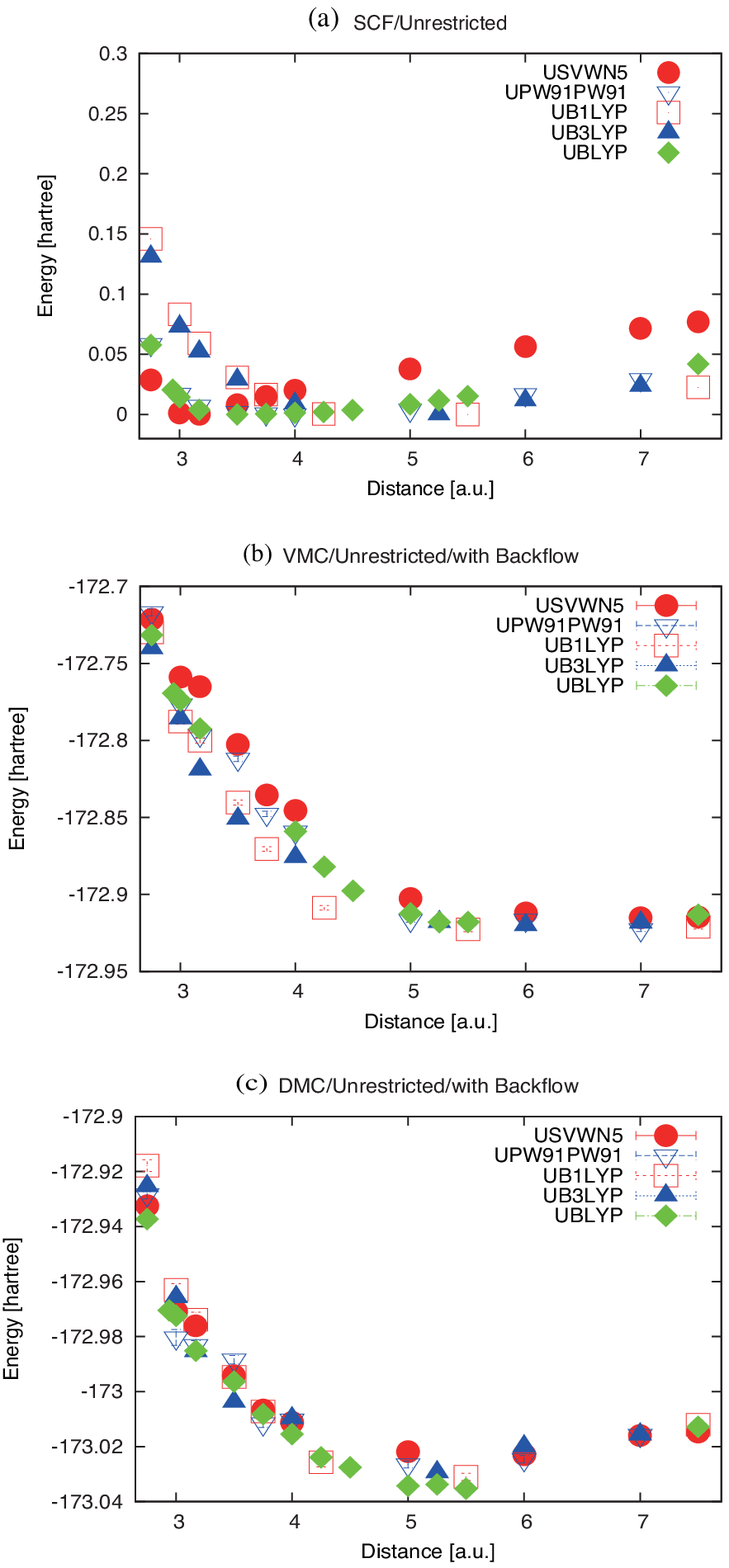}
\caption{Binding curves obtained from unrestricted (a) SCF, (b) VMC, and (c) DMC calculations.
In (a), SCF energies are shifted so that each minimum is set to be zero; the minimums are, $-172.758, -173.208, -173.026, -172.946$, and $-173.026$ hartree for SVWN5, PW91PW91, BLYP, B1LYP, B3LYP, respectively. In (b) and (c) error bars are within symbol size.
VMC and DMC include the backflow transformation. 
}
\label{SdU}
\end{figure}

Binding curves obtained from SCF and QMC calculations using unrestricted SD wavefunctions 
are shown in Fig. \ref{SdU} (a) - (c).
At the SCF level, LDA (USVWN5) and GGA(UPW91PW91 and UBLYP) 
recover a bond length, $R_e$, 
near to the experimental value, 3.1748 [a.u.]\cite{Furche_Perdew_2006}, 
while the other hybrid functionals give a longer $R_e$.
At the QMC level, using any of the XC functionals, however, the bond length, $R_e$, 
is very similar and much overestimated compared to experiment.
The results imply that the non-local HF exchange favors a longer $R_e$. 
QMC essentially takes the non-local exchange into account 
even if the trial/guiding wavefunction is constructed without the HF exchange.
VMC gives a quite shallow bottom of the binding curve, while DMC gets the curve recovered deeper. However, DMC still unsatisfactorily recovers about  30$\%$ of the experimental binding energy $D_e$, 0.0541 hartree\cite{Furche_Perdew_2006}.

\par
At the experimental $R_e$ some of the SCF calculations give lower molecular energies 
than twice the isolated atomic energies
(zero-binding energies shown in Table \ref{ZeroBindingRyo}),
as shown in Table \ref{EnergyAtExp}.
LDA(USVWN5) and GGA(UPW91PW91) recover
164.5 \% and 70.3 \% of the experimental $D_e$, respectively, 
while UB3LYP gives 3.7 \%.
Underlined values  in 
Table \ref{EnergyAtExp}
highlight 
those lower than the zero-binding energies, which means that 
the molecule is bound at the experimental $R_e$.
For example, RB3LYP-SCF does not bind the molecule, which is consistent
with previous works \cite{Edgecombe_Becke_1995,Kitagawa_etal_2007}. 
To compare the atomic and molecular energies, 
the same basis sets and pseudopotentials were used.
The RHF, UHF, and QMC calculations using the RHF and UHF trial wavefunctions 
were obtained by GAMESS/QWalk
while the others are by Gaussian03/CASINO.
A lower DFT-SCF energy does not mean a better description of the total energy
because it does not necessarily satisfy the variational principle,
while a lower QMC energy implies a solution closer to the exact one 
because of its variational property of the total energy.
In Table \ref{EnergyAtExp}, therefore, only the best (lowest) zero-binding energies 
of VMC and DMC using the B1LYP trial/guiding wavefunction are shown.
Hence none of the QMC calculations at the SD level can reproduce a bound state at the experimental $R_e$.

\par
It is worth noting in Table \ref{EnergyAtExp} that
RHF gives a lower DMC energy than UHF.
At the SCF level, in turn, RHF gives a much higher energy than any other,
supporting the consensus that restricted treatments cannot
well describe the localized amplitude of the wavefunction
at ionic sites of a spin polarized system \cite{SZA89}.
In DMC, however, the amplitude of the many-body wavefunction
is automatically adjusted by the projection operation
and is not directly governed by the XC approximation.
The results show that the RHF nodal surface is superior to
the UHF one, leading us to examine restricted methods 
for the generation of the trial nodal surface.

\par
\begin{figure}[h]
\centering
\includegraphics[width=3truein,clip]{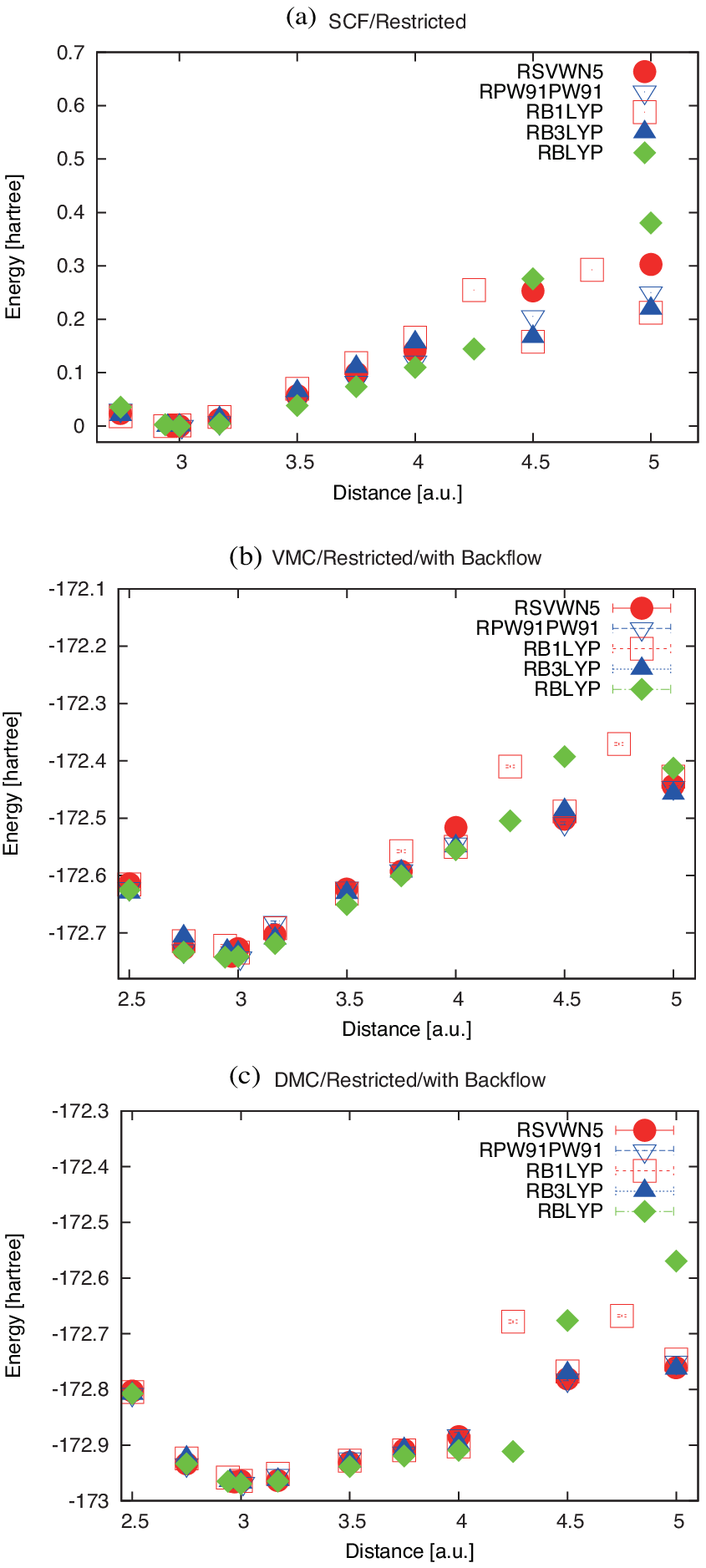}
\caption{Binding curves obtained from restricted (a) SCF, (b) VMC, and (c) DMC calculations. In (a), SCF energies are shifted so that each minimum is set to be zero; the minimums are, $-172.753, -173.180, -173.004, -172.794$, and $-172.900$ hartree for SVWN5, PW91PW91, BLYP, B1LYP, B3LYP, respectively. In (b) and (c) error bars within symbol size. VMC and DMC include the backflow transformation. }
\label{SdR}
\end{figure}

Figure \ref{SdR} (a) - (c) shows binding curves
evaluated by restricted methods.
At the SCF level, they well reproduce the experimental $R_e$.
A quite deep well bottom gives rise to an overestimate of $D_e$, which is
attributed to a well-known failure of restricted methods~\cite{SZA89}.
Such an overestimate is modestly improved in DMC because the imaginary projection relaxes
the many-body wavefunction at larger distances. 
From the viewpoint of evaluating $R_e$, RB3LYP gives the best restricted fixed nodes, 
though it is found to be variationally worse than any of unrestricted nodes.
The variationally optimal fixed nodes within the SD approximation are obtained from UB3LYP, 
though it overestimates $R_e$ by around 50 \% and
underestimates $D_e$ by 30 \%. 
The backflow transformation improves the ground state energy,
but makes no significant improvements on $R_e$ and $D_e$, 
as shown in Fig. \ref{CompBf} (a) and (b).

\par
\begin{figure}[h]
\centering
\includegraphics[width=3truein,clip]{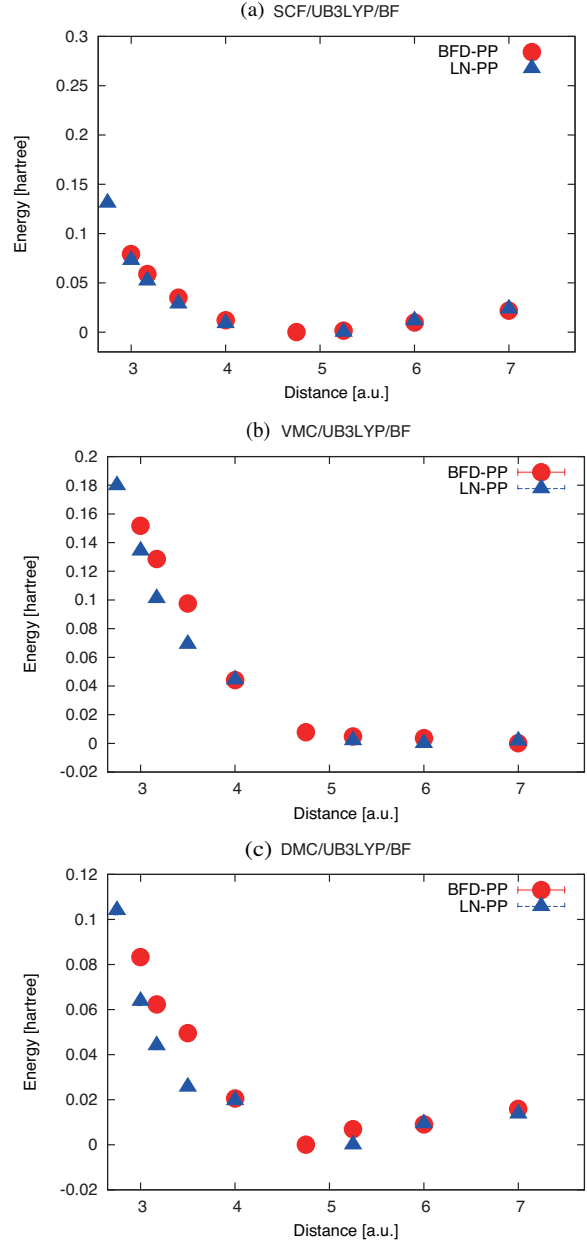}
\caption{Comparison of binding curves using different pseudopotentials with UB3LYP/BF for (a) SCF, (b) VMC, and (c) DMC calculations.
Energies
are shifted so that each minimum to be zero: the values of the minimum are (a) $-173.026$ hartree for LN and $-173.849$ hartree for BFD, (b) $-172.920$ hartree for LN and $-173.746$ hartree for BFD, and (c) $-173.029$ hartree for LN and $-173.844$ hartree for BFD. }
\label{PPcomp}
\end{figure}

We also investigated the dependence on 
pseudopotentials for the best fixed nodes within the SD approximation, 
{\it i.e.}, we performed UB3LYP-QMC with backflow 
for two different potentials, LN and BFD. 
The latter uses a contracted Gaussian basis set 
(33s29p19d2f)/ $\left[ {\rm 5s5p4d2f } \right]$\cite{BUR08}.
Though both pseudopotentials have the same effective valence electrons,
the atomic energies are quite different from each other, as shown in
Table \ref{PpZeroBindingEnergy} in terms of zero-binding energies.
We note that considerably faster SCF convergence is observed with BFD, 
by a factor of five faster than LN, 
while the QMC statistical qualities (statistical noise, auto-correlation, and population fluctuation) 
are almost the same for both potentials.
The LN/UB3LYP-SCF calculation gives a weak binding at the experimental $R_e$, 
but such a bound state disappears for BFD/UB3LYP-SCF (unbound), 
as seen in Table \ref{PpZeroBindingEnergy}.
Comparisons of binding curves are shown in Fig. \ref{PPcomp} (a) - (c).
There is no significant difference between LN and BFD in the predictions of $R_e$ and $D_e$,
though BFD slightly underestimates $D_e$ and overestimates $R_e$.
In summary, this result may justify our choice of the LN pseudopotential.

\par
\begin{figure}[h]
\centering
\includegraphics[width=3truein,clip]{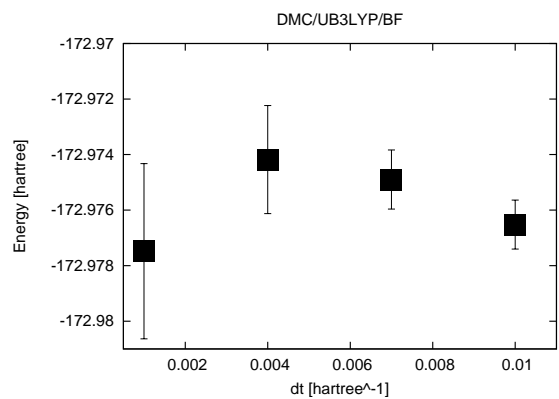}
\caption{Time step dependence of ground state energy in the DMC calculation 
with the backflow transformation. The fixed node is generated from the UB3LYP-DFT calculation. }
\label{TimeStepBias}
\end{figure}

The time step dependence of the DMC energies is shown in Fig. \ref{TimeStepBias}.
It is confirmed that the result with $dt=0.01$ agrees with the results using other choices of $dt$ 
to within one standard deviation $\sigma$.
The sudden decrease in the energy at $dt=0.001$ is found to be similar to Fig. 7 of Ref. \cite{UMR93}.

\subsection{Restricted multi determinant calculations}
\label{Restricted multi determinant calculations}
Valence bond (VB) type many-body wavefunctions are expected to 
give a proper description of the spin polarized Cr$_2$\cite{BAU80a}
with a compact form of the MD expansion. 
Generalized VB (GVB) SCF is available in GAMESS\cite{GAM93}
and we can use GVB orbitals to generate QMC trial/guiding wavefunctions.
Table \ref{NoLevel} shows the symmetries of the UHF natural orbitals (NO)
near the HOMO-LUMO level.
We considered 12 active occupied molecular orbitals up to level 20, 
which arises from the 4s(1)3d(5) atomic orbitals. 
The GVB function is formed as
\begin{equation}
\label{(10.3.17.1)}
\Psi _{\rm GVB}^{\left( 6 \right)}  = {\hat A}\left\{ {\Phi _{\rm core}
^{\left( 8 \right)} \cdot\prod\limits_{p = 1}^6 
{\Phi _{{\rm GVB};p}^{\left( 2 \right)} } } \right\} \ ,
\end{equation}
where $\hat A$ denotes an antisymmetrizer and $\Phi _{\rm core}^{\left( 8 \right)}$ the core contribution.
$\Phi _{{\rm GVB};p}^{\left( 2 \right)}$ consists of the (GVB) orbital pairs,
for which each of $j$-occupied orbitals ($j \in \left\{ {9, \cdots ,14}\right\}$)
in the active space is paired with 
a virtual orbital 
with the same
symmetry. Hence the following six pairs are involved:
$p=(10,19),(11,18),(14,15),(13,16),(12,17)$, and $(9,20)$.
The orbitals in the active space were optimized by a SCF procedure with respect to 
$\Psi _{\rm GVB}^{\left( 6 \right)}$.
An explicit form of $\Psi _{\rm GVB}^{\left( 6 \right)}$ takes  
a restricted CI expansion, $D_j^\uparrow=D_j^\downarrow$, $c_{ij} = c_i \cdot \delta_{ij}$,
as,
\begin{equation}
\label{(10.3.17.2)}
F_{\rm AS}^{\rm GVB}
=\sum_{i=0}^{63}{c_i\cdot D_i^\sigma}
= \left[ \begin{array}{l}
 \left( {g_9 \cdot\hat 1 + e_9 \cdot\hat P_{9 \to S_9 } } \right) \\ 
  \otimes \left( {g_{10} \cdot\hat 1 + e_{10} \cdot\hat P_{10 \to S_{10} } } \right) 
\\ 
  \otimes  \cdots  \\ 
  \otimes \left( {g_{14} \cdot\hat 1 + e_{14}\cdot\hat P_{14 \to S_{14} } } \right) 
\\ 
 \end{array} \right]\cdot D_0^\sigma \ ,
\end{equation}
where $g_j$ and $e_j$ are coefficients such that
$c_0 = g_9\cdot g_{10} \cdots g_{14}\,$ 
{\it etc}.
$S_j$ denotes the index of 
a virtual orbital
with the same symmetry
as the $j$-occupied orbital.
The operators, $\hat 1$ and $\hat P_{j \to S_j }$,
are the identity and permutation operators, respectively.  
$\hat P_{j \to S_j }$ swaps the $j$-occupied orbital in $D_0^\sigma $
into the 
$S_j$-virtual one.
As discussed later a usual CI treatment in a quantum chemistry code
gives hundreds of thousands of terms in the expansion, all of which can not be included 
in a QMC calculation.
GVB provides, in contrast, a very compact form of the MD expansion with
only 64 terms.

\par
Starting with UHF-NO and optimizing the coefficients, $g_j$ and $e_j$, as well as the orbitals, 
GVB-SCF gives a variationally better description than RHF
within a restricted SCF treatment, as seen in Table \ref{GVBresults}.
With the coefficients given by GVB-SCF, the wavefunction achieves a better (lower)
energy than HF at the VMC level, but it turns out to be higher than HF using DMC
(the row of 'GVB(6)' in Table \ref{GVBresults}).
Then we tried to optimize the coefficients further by VMC.
A total of 64 coefficients can be reduced to 48 independent variables by its symmetry.
We adopted a mixed scheme between energy and
variance minimization \cite{UMR05} with 95\% weight on the former.
Even though we ignore the above symmetry reduction to optimize 64 parameters
independently, the optimized values of the parameters roughly satisfy their symmetries.
Using the GVB nodes with these coefficients, we obtained a better DMC value
than when using the HF nodes, but still above the zero-binding energy
(the row of 'GVB(6)opt' in Table \ref{GVBresults}).

\par
\begin{figure}[h]
\centering
\includegraphics[width=3truein,clip]{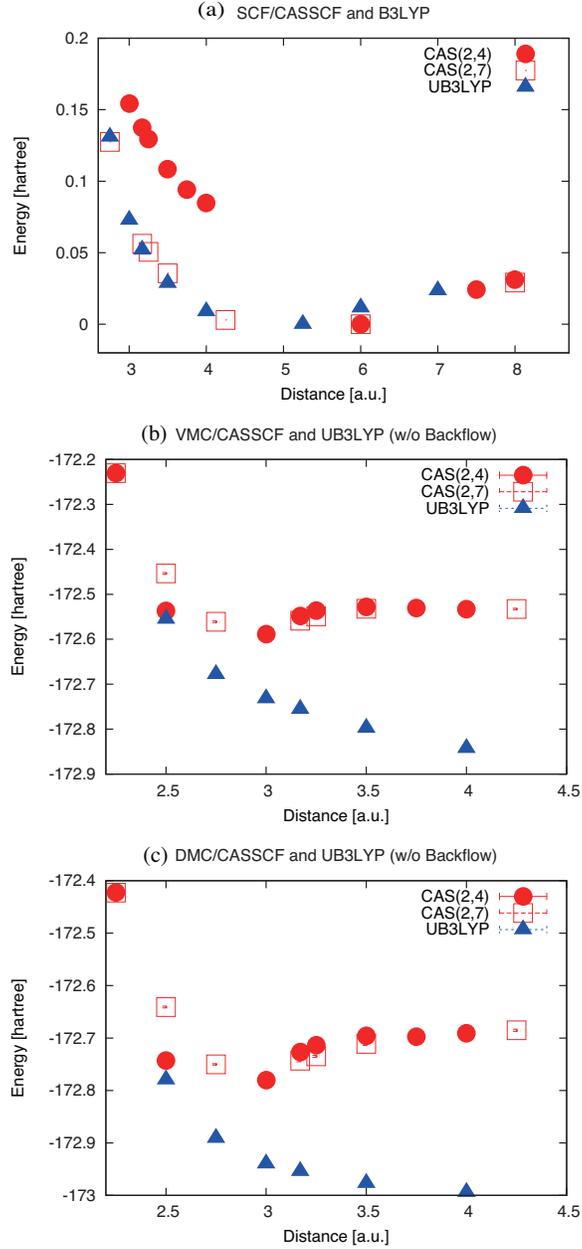}
\caption{(a) SCF, (b) VMC, and (c) DMC binding curves using the CASSCF(2,4), CASSCF(2,7), and UB3LYP (trial) wavefunctions. In (a) SCF energies are shifted so that each minimum, $-171.540 , -171.461$, and $-173.026$ hartree to be zero. In (b) and (c) error bars are within symbol size. }
\label{casscf}
\end{figure}

Next we considered a restricted CASSCF node ($D_j^\uparrow=D_j^\downarrow$,
$c_{ij} = c_i \cdot \delta_{ij}$), having the form of 
\begin{equation}
\label{(10.3.17.2)}
F_{\rm AS}^{\rm CAS}
=\sum_{i=0}^{N_{\rm CAS}-1}{c_i\cdot D_i^\sigma}
=\sum_{i=0}^{N_{\rm CAS}-1}{c_i\cdot \hat P_{l \rightarrow m}^{\left(i\right)}
D_0^\sigma} \ .
\end{equation}
An initial guess for the orbitals was taken from a UHF-NO calculation and
optimized self-consistently with respect to the above many-body wavefunction.
We employed (restricted) CASSCF(2,4) and CASSCF(2,7) methods, in which
the number of expansion terms amounts to $N_{\rm CAS}$=16 and 49, respectively.
The results are shown in Fig. \ref{casscf} (a) - (c).
Though they could not achieve a variationally lower energy than 
UB3LYP-SD in terms of the final DMC energy, several interesting behaviors are found as follows:
CASSCF(2,7)-SCF gives a binding curve with a similar shape to
UB3LYP-SCF, though overestimating $R_e$.
At the QMC level, in turn, the evaluated value of $R_e$ gets shorter,
and the shape of the binding curve is similar to the restricted SD cases.
This implies that the terms in the MD expansion well describe the 
localized amplitude as that obtained from the unrestricted SD cases,
but the nodal structure is essentially the same as that obtained from the restricted SD.
In order to obtain a variationally lower energy we have tried a 
restricted MRCI (multi reference CI) using orbitals obtained from
the present CASSCF calculation, but QMC calculations using the MRCI trial/guiding wavefunction
could not give better results than when using the UB3LYP-SD one.

\subsection{Unrestricted multi determinant calculations}
Several advanced MD implementations such as CASSCF 
are available at the restricted level, but we could not obtain
better results than UB3LYP-SD.
We therefore tried unrestricted CI (UCI) methods, for which 
Gaussian03~\cite{Gaussian03} was used to provide UCISD (UCI singles and doubles).
Using a UHF reference, the UCI expansion gives 7,521,823 terms,
all of which can not  be taken into account in a QMC calculation.
We truncated this expression into 35 determinants, 
removing those terms with coefficients $\left| c_i \right| < 0.01 $.
The expansion coefficients were optimized further by VMC,
first by weight-limitted variance minimizations \cite{CASINO}, 
followed by energy minimizations.
The results at the experimental $R_e$ are shown in Table \ref{UCISD-SCF}.
At the SCF level, UCISD achieved a lower energy than its initial guess UHF,
implying that the coefficients are well optimized by CI-SCF.
UCISD gives a higher energy than UB3LYP at the SCF level, which does not matter, 
because there is no variational relation between them.
Using the UCISD trial wavefunction, we achieved a lower VMC energy 
than when using the UB3LYP one, 
indicating that our VMC optimization
of the CI coefficients was successful.
At DMC, however, 
UCISD turned out to give a worse result than UB3LYP.
We also tried another choice of expansion: 67 excited configurations, 
in which the active space was HOMO$\pm$6 and 
the single and double excitations of occupied orbitals were restricted 
to 
virtual
orbitals with the same symmetry.
This choice gives, however, a worse result than UB3LYP-SD even at the
VMC level [-172.744(2) hartree].

\par
The above CI treatments could not give any variationally better trial node than UB3LYP-SD.
The easiest way to go beyond those treatments would be to add UCI expansions
to UB3LYP-SD because it is the best starting point.
By considering the UB3LYP orbital symmetry near the HOMO level,
we made two different sizes of CI expansions:
The first one took into account only 3 virtual orbitals
above the LUMO, including only $\sigma$ and $\pi$ symmetries
[for which we refer it as 
UCISD(UB3LYP+3)], and the second was
a larger one with 10 virtual orbitals in which
$\sigma$, $\pi$, and $\delta$ symmetries were included 
[UCISD(UB3LYP+10)].
In both cases we considered only such excited configurations
between the orbitals with the same symmetry, resulting in
around 50 and 650 determinants for
UCISD(UB3LYP+3)
and  
UCISD(UB3LYP+10),
respectively (the numbers of determinants
vary a little amount depending on $R$). 
\par
\begin{figure}[h]
\centering
\includegraphics[width=3truein,clip]{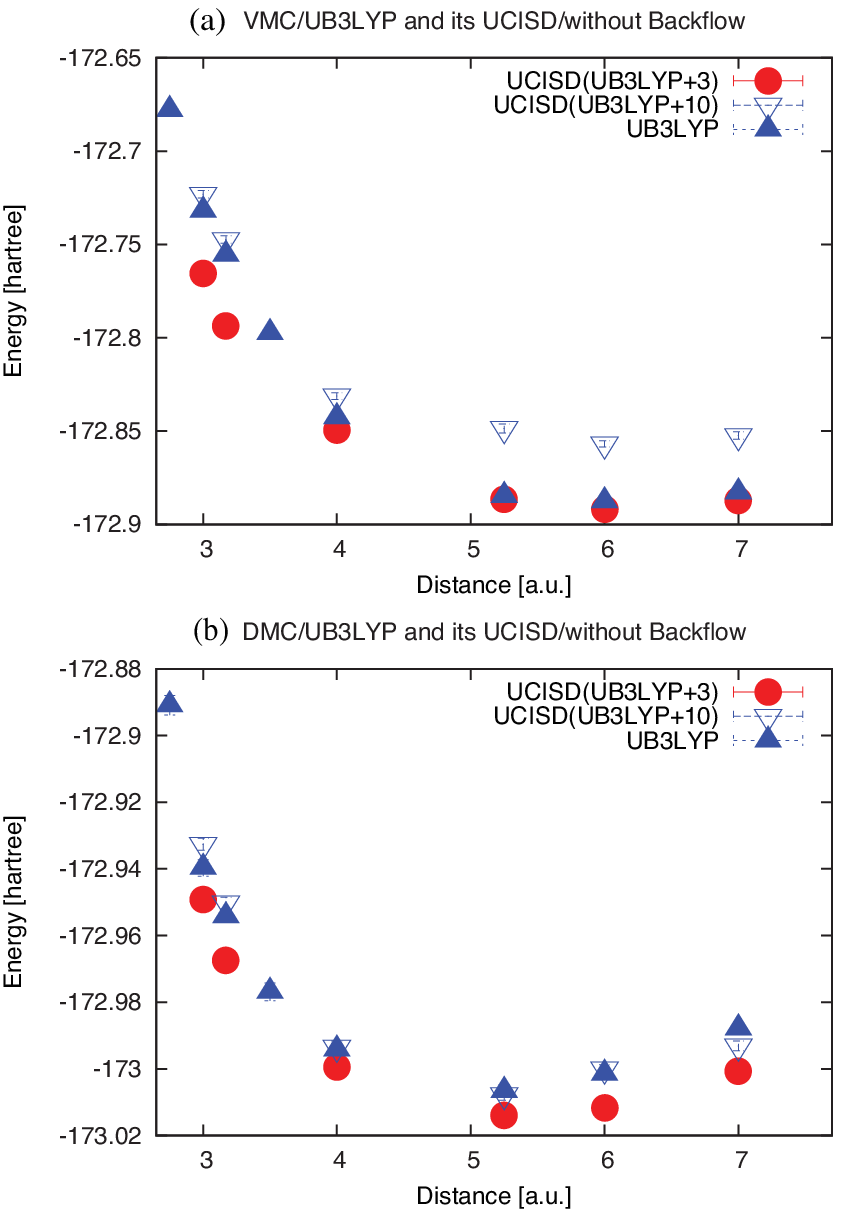}
\caption{Comparison of QMC binding curves using the UB3LYP-SD and 
UCISD(UB3LYP+$N$) ($N = 3, 10$)
trial/guiding wavefunctions for (a) VMC and (b) DMC without Backflow transformation.}
\label{UCISD/woBF}
\end{figure}
\begin{figure}[h]
\centering
\includegraphics[width=3truein,clip]{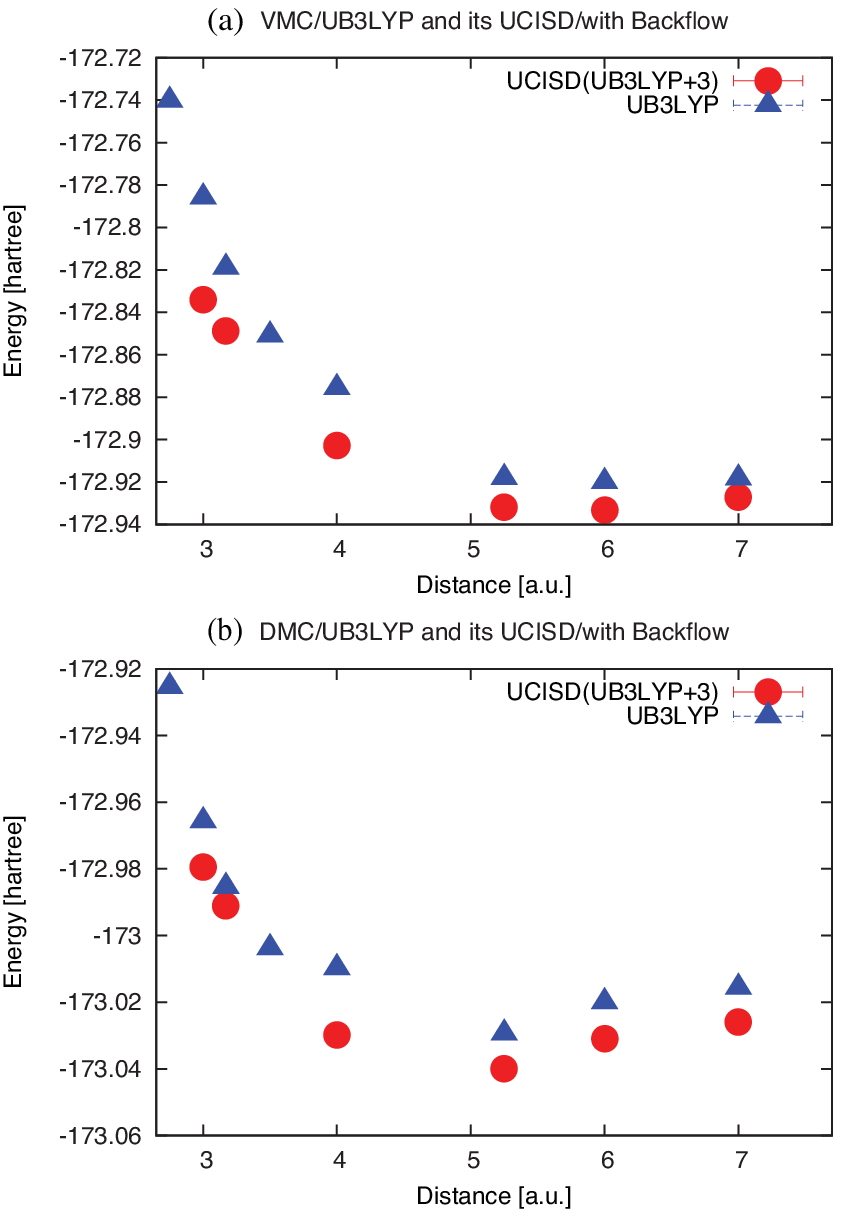}
\caption{Comparison of QMC binding curves using the UB3LYP-SD and 
UCISD(UB3LYP+3)
trial/guiding wavefunctions for (a) VMC and (b) DMC with the backflow transformation.}
\label{UCISD/withBF}
\end{figure}
\begin{figure}[h]
\centering
\includegraphics[width=3truein,clip]{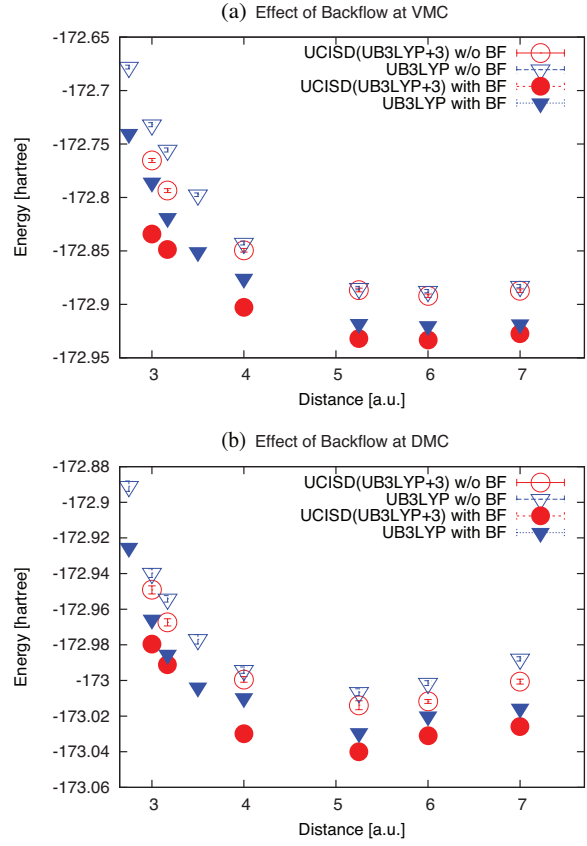}
\caption{Comparison between results with and without the
backflow transformation for (a) VMC and (b) DMC.}
\label{CompBf}
\end{figure}

The results are shown in 
Fig. \ref{UCISD/woBF} (a) and (b).
As seen in Fig. \ref{UCISD/woBF} (a), VMC using the 
UCISD(UB3LYP+3)
trial wavefunction 
gives a better result than when using the UB3LYP-SD one,
 because the former includes more variational degrees of freedom to be optimized.
For 
UCISD(UB3LYP+10),
however, we could not get a satisfactory optimization, 
giving a higher energy than the initial UB3LYP-SD calculation.
Nevertheless the DMC calculation with the 
UCISD(UB3LYP+10)
node
gives a slightly lower energy than that with the UB3LYP node (Fig. \ref{UCISD/woBF} (b)). 
It is apparent that the 
UCISD(UB3LYP+3)
node give a lower DMC energy than the UB3LYP-SD node.
Focussing on 
UCISD(UB3LYP+3),
we further
introduced the backflow transformation, getting the best binding curve
beyond SD, as shown in Fig. \ref{UCISD/withBF} (a) and (b).
Similar to the SD cases, the DMC projection gets $R_e$ shorter
and $D_e$ deeper than VMC.
Though 
UCISD(UB3LYP+3)
gives a variationally better result than any SD treatment, 
it could not hardly improve the binding nature, {\it i.e.}, it overestimates $R_e$ 
and underestimates $D_e$, compared with the experimental values.
As seen in Fig. \ref{CompBf} (a) and (b),
the backflow does not improves these as well.

\section{Concluding Remarks} 
\label{section:concluding_remarks}
\par
We studied the binding curve of the ground state Cr$_2$ dimer using the fixed node DMC method. 
Various different types of nodal structures were 
compared based on the variational principle 
with respect to the node of the DMC guiding function.
We tested several choices of XC functionals with or without
spin restriction on the orbital functions composing the many-body wavefunction.
We also tried computationally expensive choices of UCI expansions with
backflow transformation.
\par
Within the SD treatment, UB3LYP turns out to give the variationally best trial node.
Except HF, the unrestricted nodes are found to be better than the restricted ones.
Any choice gives binding curves with a energy minimum,
but for unrestricted trial nodes they end up with a much larger $R_e$ and 
smaller $D_e$ in the QMC final results, compared with the experimental values.
Though some unrestricted DFT-SCF  calculations, 
such as ULDA and UGGA, reproduce a proper $R_e$,
it is found that the QMC calculations with these trial nodes overestimate $R_e$.
The restricted nodes recover fairly well $R_e$ even at QMC,
but they give a higher energy than the unrestricted ones.
At the experimental $R_e$, we could not get a stable molecular energy 
lower than twice the atomic energy at the QMC level, 
although some DFT-SCF calculations did reproduce a bound state.

\par
We also examined whether the binding curve could be improved by different
pseudopotentials, comparing the BFD potential with the 
LN
one.
Both potentials give almost the same binding curve,
which justifies our choice of pseudopotentials. 
The time step bias is confirmed to be kept within the error bar considered in this study.
The backflow transformation turned out to give no specific
improvements on describing the binding nature, although it 
did
improve the energy
variationally.

\par
Within the framework of the MD approximation, we first tried the GVB model as a compact expansion 
for the many-body wavefunction because of its plausible physical meaning. 
Starting with HF orbitals, we optimized the GVB orbitals and coefficients by a SCF procedure,
but it could not give a better result than the unrestricted SD in the DMC final results, 
probably because the restricted treatment has a limitation 
in describing the spin polarized nature in Cr$_2$.
We also run DMC using the trial nodes derived from a restricted MD expansion 
with orbitals optimized by CASSCF,
but it ends up with a worse QMC result than the unrestricted DFT-SD nodes.
Though the restricted CASSCF itself gives a too large $R_e$ similar to the unrestricted DFT calculations,
the QMC calculations with the CASSCF nodes give properly shorten $R_e$
 near to the experimental value.
Then we decided to use the unrestricted CI nodes.
First, we chose such excited configurations 
as those with a large coefficient weight and re-optimized the coefficients. 
VMC with this trial wavefunction gives a lower energy than that with the UB3LYP one. 
On the other hand, the corresponding DMC calculation gives a higher energy than 
DMC using the UB3LYP nodes.
Next, we manually constructed a UCI expansion with UB3LYP orbitals and optimized their coefficients. 
Though the UCI-MD node gives a better DMC result than the UB3LYP-SD one,
it gives no improvements on the binding natures, {\it i.e.}, it still overestimates $R_e$ and underestimate $D_e$.

\section{Acknowledgments}
The authors would like to thank Prof. Lubos Mitas for
introducing this work to us and for providing basis sets and
pseudopotentials, Dr. Lucas Wagner for his help about the QWalk code,
Ms. Kaoru Nishi for her work on the numerical data, and Dr. Mark A. Watson for his fruitful comments.
The computation in this work has been partially performed using
the facilities of the Center for Information Science in JAIST.
K.H. is grateful for a JSPS Postdoctoral Fellowship for Research Abroad.
Financial support was provided by
Precursory Research for Embryonic Science and Technology, 
Japan Science and Technology Agency (PRESTO-JST), and
by a Grant--in--Aid for Scientific Research in Priority Areas 
"Development of New Quantum Simulators and Quantum Design (No. 17064016)''
(Japanese Ministry of Education,Culture, Sports, Science, and Technology ; KAKENHI-MEXT) for R.M.

\appendix
\section{Jastrow functions}
In CASINO \cite{CASINO}, $u$, $\chi$, and $f$ terms 
in Eq. (\ref{10.02.04.math02}) are given 
in power expansion form as \cite{DRU04} 
\begin{eqnarray}
 u\left( {r_{ij} } \right) &=& \left( {r_{ij}  - L_u } \right)^C  \times \Theta \left( {L_u  - r_{ij} } \right) \nonumber \\
 && \times \left( {\alpha _0  + \left[ {\frac{{\Gamma _{ij} }}{{\left( { - L_u } \right)^C }} + \frac{{\alpha _0 C}}{{L_u }}} \right]r_{ij}  + \sum\limits_{l = 2}^{N_u } {\alpha _l } r_{ij}^l } \right) \label{jastrowN_u} \ , \\
 \chi _I \left( {r_{iI} } \right) &=& \left( {r_{iI}  - L_{\chi I} } \right)^C  \times \Theta \left( {L_{\chi I}  - r_{iI} } \right) \nonumber \\
 &&  \times \left( {\beta _{0I}  + \left[ {\frac{{ - Z_I }}{{\left( { - L_{\chi I} } \right)^C }} + \frac{{\beta _{0I} C}}{{L_{\chi I} I}}} \right]r_{iI}  + \sum\limits_{m = 2}^{N_\chi  } {\beta _{mI} } r_{_{iI} }^m } \right) \label{jastrowN_chi} \ , \\ 
 f_I \left( {r_{iI} ,r_{jI} ,r_{ij} } \right) &=& \left( {r_{iI}  - L_{fI} } \right)^C \left( {r_{jI}  - L_{fI} } \right)^C  \nonumber \\
 && \times \Theta \left( {L_{fI}  - r_{iI} } \right)\Theta \left( {L_{fI}  - r_{jI} } \right)\sum\limits_{l = 0}^{N_{fI}^{eN} } {\sum\limits_{m = 0}^{N_{fI}^{eN} } {\sum\limits_{n = 0}^{N_{fI}^{ee} } {\gamma _{lmnI} r_{iI}^l r_{jI}^m r_{ij}^n } } }  \ .
\label{jastrowN_fI}
\end{eqnarray}
In QWalk \cite{QWA09}, instead, the terms are expanded
with basis sets $\left\{ {b_j \left( r \right)} \right\}$ 
which vanish at some cutoff length 
{\color{red}
,
}
given as
\begin{eqnarray}
u\left( {r_{ij} } \right) 
&=& u_0 \left( {r_{ij} } \right) 
+ \sum\limits_{m = 1}^{M_u } {c_m^{} \cdot b_m^{\left( {ee} \right)} \left( {r_{ij} } \right)} \ ,
\\
\chi _I \left( {r_{iI} } \right) 
&=& \sum\limits_{m = 1}^{M_\chi} {c_m^{} \cdot b_m^{\left( {eI} \right)} \left( {r_{iI} } \right)} \ ,
\\ 
f_I \left( {r_{iI} ,r_{jI} ,r_{ij} } \right) 
&=& \sum\limits_{\left\langle {k,m,n} \right\rangle }^{} {c_{kmn}^{} \cdot b_m^{\left( {eI} \right)} \left( {r_{iI} } \right)b_n^{\left( {eI} \right)} \left( {r_{jI} } \right)\cdot b_k^{\left( {ee} \right)} \left( {r_{ij} } \right)} \ ,
\end{eqnarray}
with
\begin{eqnarray}
u_0 \left( r \right) 
&=& \frac{1}{4}\cdot\frac{{p_{ \uparrow  \uparrow } \left( r \right)}}{{1 + \underline {\gamma _{ \uparrow  \uparrow } } \cdot p_{ \uparrow  \uparrow } \left( r \right)}}\cdot\hat P_{ \uparrow  \uparrow }^{}  + \frac{1}{2}\cdot\frac{{p_{ \uparrow  \downarrow } \left( r \right)}}{{1 + \underline {\gamma _{ \uparrow  \downarrow } } \cdot p_{ \uparrow  \downarrow } \left( r \right)}}\cdot\hat P_{ \uparrow  \downarrow } \ ,
\\
p_{\sigma \sigma '} \left( r \right) 
&=& \left( {\frac{r}{{r_c^{\sigma \sigma '} }}} \right) - \left( {\frac{r}{{r_c^{\sigma \sigma '} }}} \right)^2  + \frac{1}{3}\left( {\frac{r}{{r_c^{\sigma \sigma '} }}} \right)^3 \ ,
\\
\label{eq_a9}
b_m^{\left( {pq} \right)} \left( r \right) 
&=& \frac{{1 - z^{\left( {pq} \right)} \left( r \right)}}{{1 + \beta _m^{\left( {pq} \right)} 
\cdot z^{\left( {pq} \right)} \left( r \right)}}\ ,
\\
\label{eq_a10}
z^{\left( {pq} \right)} \left( r \right)
&=& \left( {\frac{r}{{b_0^{\left( {pq} \right)} }}} \right)^2 \left[ {6 - 8\left( {\frac{r}{{b_0^{\left( {pq} \right)} }}} \right) + 3\left( {\frac{r}{{b_0^{\left( {pq} \right)} }}} \right)^2 } \right] \ ,
\end{eqnarray}
where the upper index $\left(pq\right)$ in Eqs. \ref{eq_a9} and \ref{eq_a10} stands for
particle pairs such as $\left(eI\right)$ or $\left(ee\right)$.
The term $u_0$ imposes the electron-electron cusp condition 
for spin pair $\sigma\sigma'$
with a projection operator $\hat P_{ \sigma  \sigma'}$ and cutoff length $r_c^{\sigma\sigma'}$.
The Poly-Pade type basis $b_m^{\left( {pq} \right)}$ with
$z^{\left( {pq} \right)} \left( r \right)$ is designed to be 
cutoff quadratically at $b_0^{\left( {pq} \right)}$.
The parameters $\left\{\beta_m^{\left( {pq} \right)}\right\}$
are generated from a given $\beta_0^{\left( {pq} \right)}$
using the following recursions:
\begin{eqnarray}
\beta_1  
&=& \exp \left( {1.6} \right) \times \left( {1 + \beta _0^{} } \right)\ ,
\\ 
\beta_k &=& \left[ {\exp \left( {1.6} \right)} \right] \times \beta _{k - 1}
\quad
\left(k>1\right)\ .
\end{eqnarray}
The range of summation for the $f$-term,
$\left\langle {k,m,n} \right\rangle$, denotes the pairs
specified according to a given order $\left(M_u,M_\chi \right)$.
For the present work $\left(M_u,M_\chi \right)=\left(3,3 \right)$ 
amounts
to 12 terms for the expansion.

For CASINO we used expansion orders of 
$N_{\rm u}$=8, $N_{\rm \chi}$=8,
$N_{\rm fI}^{eN}$=2, and $N_{\rm fI}^{ee}$=2, with fixed
cutoff lengths $L_{\rm u}$ = 5 [a.u.],
$L_{\rm \chi}$ = 4 [a.u.], and $L_{\rm fI}$ = 3[a.u.], respectively.
In QWalk we choose cutoff lengths to be fixed as 7.5 [a.u.].


\clearpage

\clearpage
\begin{table}[h]
\caption{Exchange-Correlation potentials. `GC' stands for gradient correction.}
\label{xc}
\begin{center}
\begin{tabular}{c | rrc | rc}
\noalign{\hrule height 1pt}
 & \multicolumn{3}{c|}{Exchange} & \multicolumn{2}{c}{Correlation}\\
 & Non-local & Local & GC & Local & GC\\
Functional & $V_{\rm X}^{\rm HF}$[$\%$] & $V_{\rm X}^{\rm Slater}$[$\%$]  
& $\delta V_{\rm X}$ [$\%$] &  $V_{\rm C}^{\rm VWN5}$[$\%$]  & $\delta V_{\rm C}$ [$\%$] \\
\noalign{\hrule height 1pt}
SVWN5 & 0 & 100 & 0 & 100 & 0\\
PW91PW91 & 0 & 100 & 100~\!PW91 & 100 & 100~\!PW91\\
BLYP & 0 & 100 & 100~\!B88 & 100 & 100~\!LYP\\
B1LYP & 25 & 75 & 75~\!B88 & 100 & 100~\!LYP\\
B3LYP & 20 & 80 & 72~\!B88 & 100 & 81~\!LYP\\
\noalign{\hrule height 1pt}
\end{tabular}
\end{center}
\end{table}
\begin{table}[h]
\caption{Ground state energies 
[hartree]
at the experimental bond length.
Underlined values 
highlight
those lower than the zero-binding energy in Table \ref{ZeroBindingRyo}.
}
\label{EnergyAtExp}
\begin{center}
\begin{tabular}{c | ccccc}
\noalign{\hrule height 1pt}
  &SCF & VMC & DMC   \\
\noalign{\hrule height 1pt}
UHF & -171.701& -172.665(6) &  -172.903(2) \\
RHF & -171.097  & -172.584(4) &  -172.911(2) \\
USVWN5 & \underline{-172.757} & -172.765(2) & -172.976(2) \\
RSVWN5 & \underline{-172.741} & -172.704(2) &  -172.963(3) \\
UPW91PW91 & \underline{-173.201} & -172.796(2) &  -172.983(2)  \\
RPW91PW91 & \underline{-173.173} & -172.681(2) & -172.954(2)  \\
UBLYP &  \underline{-173.022}   &  -172.792(2)   &  -172.985(2)\\
RBLYP &    \underline{-173.000} &   -172.719(2) &   172.965(1)  \\
UB1LYP & -172.887 & -172.800(2) & -172.974(3) \\
RB1LYP & -172.777 & -172.692(2) &-172.952(2) \\
UB3LYP(without BF)& \underline{-172.974} & -172.756(2) &  -172.954(2) \\
UB3LYP(with BF)& \underline{-172.974} & -172.819(2) & -172.985(2) \\
RB3LYP & -172.885 & -172.710(2) & -172.963(2) \\
\hline
ZeroBinding(Best) &&-172.931(2)&-173.012(3)\\
\noalign{\hrule height 1pt}
\end{tabular}
\end{center}
\end{table}
%

\begin{table}[h]
\caption{Zero-binding energies 
[hartree]
for different trial wavefunctions.
QMC values are evaluated with the backflow transformation.}
\label{ZeroBindingRyo}
\begin{center}
\begin{tabular}{c | ccccc}
\noalign{\hrule height 1pt}
  &SCF & VMC & DMC   \\
\noalign{\hrule height 1pt}
SVWN5 & -172.668 & -172.918(2) & -173.007(1)  \\
PW91PW91 & -173.163 & -172.919(2) & -173.007(1)  \\
BLYP &  -172.972 &-172.925(2) &  -173.008(2)  \\
B1LYP & -172.913 & -172.931(2) & -173.012(3)   \\
B3LYP & -172.972 & -172.910(2) & -173.000(2) \\
HF &-171.875 & -172.921(2)&  -173.002(2) \\
\noalign{\hrule height 1pt}
\end{tabular}
\end{center}
\end{table}
%
\begin{table}[h]
\caption{Comparison of zero-binding energies 
[hartree]
for different pseudopotentials. 
SCF molecular energies are also listed for comparison,
designated as SCF (molecule).}
\label{PpZeroBindingEnergy}
\begin{center}
\begin{tabular}{c | cccc}
\noalign{\hrule height 1pt}
pseudopotentials & SCF & SCF (molecule) & VMC & DMC \\
\noalign{\hrule height 1pt}
Lee-Needs (LN) & -172.972 & -172.974 & -172.910(2) &  -173.000(2) \\
Burkatzki-Filippi-Dolg (BFD) & -173.809 & -173.780 & -173.740(2) &  -173.820(2) \\
\noalign{\hrule height 1pt}
\end{tabular}
\end{center}
\end{table}
%
\begin{table}[h]
\caption{Symmetries of the UHF natural orbitals near the HOMO (14) and 
LUMO (15) levels.} 
\label{NoLevel}
\begin{center}
\begin{tabular}{c | c}
\noalign{\hrule height 1pt}
Level & Symmetry  \\
\noalign{\hrule height 1pt}
20	&	$\sigma_{4s}$ \\
19	&	$xz$	\\
18	&	$yz$	\\
17	&	$\sigma_{z^2}$	\\
16	&	$xy$	\\
15	&	$x^2-y^2$ \\	
14	&	$x^2-y^2$ \\	
13	&	$xy$	\\
12	&	$\sigma_{z^2}$	\\
11	&	$yz$	\\
10	&	$xz$	\\
9	&	$\sigma_{4s}$	\\
\noalign{\hrule height 1pt}
\end{tabular}
\end{center}
\end{table}
%
\begin{table}[h]
\caption{GVB  
energies [hartree]
at experimental bond length.
`GVB(6)opt' stands for that with coefficients optimized further by VMC (see text).
}
\label{GVBresults}
\begin{center}
\begin{tabular}{c | ccccc}
\noalign{\hrule height 1pt}
Methods & SCF & VMC  & DMC \\
\noalign{\hrule height 1pt}
UHF & -171.701& -172.665(6) &  -172.903(2) \\
RHF & -171.097  & -172.584(4) &  -172.911(2) \\
GVB(6) & -171.624 & -172.698(6) & -172.861(2) \\
GVB(6)opt & & -172.723(2)& -172.933(2) \\
\noalign{\hrule height 1pt}
\end{tabular}
\end{center}
\end{table}
%
\begin{table}[h]
\caption{Comparison between 
UCISD-MD and UB3LYP-SD 
at experimental bond length. QMC results are 
those without the backflow transformation.
Underlined value
highlights
that reproducing the binding.}
\label{UCISD-SCF}
\begin{center}
\begin{tabular}{c | ccccc}
\noalign{\hrule height 1pt}
Methods & SCF & VMC  & DMC \\
\noalign{\hrule height 1pt}
UHF & -171.701& -172.665(6) &  -172.903(2) \\
UB3LYP(w/o BF) & \underline{-172.974} & -172.756(2) &  -172.954(2) \\
UCISD(w/o BF) & -172.599 & -172.772(2) & -172.926(3) \\
\hline
ZeroBinding(Best) &&-172.931(2)&-173.012(3)\\
\noalign{\hrule height 1pt}
\end{tabular}
\end{center}
\end{table}


\end{document}